\documentclass[epj]{svjour}
\usepackage{graphicx}
\usepackage{amssymb}

\begin{document}

\title{Engineering decoherence in Josephson persistent-current qubits} 
\subtitle{Measurement apparatus and other electromagnetic environments}
\author{Caspar H. van der Wal
\thanks{Present address: Department of Physics, Harvard University, 17 Oxford Street, Cambridge, MA 02138, USA. E-mail: cvanderwal@cfa.harvard.edu},
F. K. Wilhelm
\thanks{Present address: Sektion Physik and CeNS, Ludwig-Maximilians Unversit\"{a}t, Theresienstr. 37, 80333 Munich, Germany},
C. J. P. M. Harmans, \and J. E. Mooij}
\institute{Department of Applied Physics and Delft Institute for Micro Electronics and Submicron Technology (DIMES),\\
Delft University of Technology, P.~O.~Box 5046, 2600 GA Delft, the Netherlands}

\abstract{We discuss the relaxation and dephasing rates that result from the
control and the measurement setup itself in experiments on Josephson
persistent-current qubits. For control and measurement of the
qubit state, the qubit is inductively coupled to electromagnetic
circuitry. We show how this system can be mapped on the spin-boson
model, and how the spectral density of the bosonic bath can be
derived from the electromagnetic impedance that is coupled to the
qubit. Part of the electromagnetic environment is a measurement
apparatus (DC-SQUID), that is permanently coupled to the single
quantum system that is studied. Since there is an obvious conflict
between long coherence times and an efficient measurement scheme,
the measurement process is analyzed in detail for different
measurement schemes. We show, that the coupling of the measurement 
apparatus to the qubit can be controlled {\em in situ}. 
Parameters that can be realized in
experiments today are used for a quantitative evaluation, and it
is shown that the relaxation and dephasing rates that are induced
by the measurement setup can be made low enough for a
time-resolved study of the quantum dynamics of Josephson
persistent-current qubits. Our results can be generalized as engineering
rules for the read-out of related qubit systems.
}
\authorrunning{Caspar H. van der Wal \textit{et al.}}
\titlerunning{Engineering decoherence in Josephson persistent-current qubits}
\maketitle


\section{Introduction}

The dynamics of electromagnetic circuits and other macroscopic objects is 
usually well described by classical laws; quantum coherent phenomena like 
superposition states are usually not observed in macroscopic systems.
The founders of the quantum mechanical theory already recognized that there
is in fact a conflict between a straight forward extrapolation of quantum 
mechanics to a macroscopic scale, and the laws of classical physics 
that govern the macroscopic world. In particular, this concerns the possibility 
of quantum superpositions of collective coordinates (i.~e.~center-of-mass-like
coordinates) of objects that are much bigger than the atomic scale. These
difficulties were first presented by Schr\"{o}dinger \cite{schrodinger35},
and are now known as Schr\"{o}dinger's cat paradox. Schr\"{o}dinger's
discussion of the cat in the box was clearly meant as a {\it gedanken}
experiment. Only several decades later, after the discovery of the Josephson
effect, it was recognized that the validity of quantum mechanics for a
macroscopic degree of freedom could be tested in {\it real} experiments 
\cite{anderson64}.

In 1980, Leggett pointed out that cryogenic and microfabrication
technologies had advanced to a level where macroscopic Schr\"{o}dinger's cat
states could possibly be realized in small superconducting loops that
contain Josephson tunnel junctions \cite{leggett80,leggett85}. In such systems, the
Josephson phase (or equivalently, the persistent supercurrent in the loop)
is a collective coordinate for the Cooper-pair condensate, and it is
conjugate to a variable which describes the charge difference across the
Josephson tunnel junction. However, while the analysis of the isolated
quantum system shows that superpositions of the macroscopic coordinates
might very well occur in these loops, it is by no means obvious that such
behavior can also be demonstrated experimentally. 
Such superposition states are extremely fragile, reflecting the 
tendency of macroscopic systems towards classical behavior.
Besides decoherence from a weak coupling to
the environmental degrees of freedom inside the solid-state
device (which is believed to be very much suppressed at low temperatures
due to the energy gap for quasiparticle excitations in superconductors), also the fact that 
the loop is not isolated but permanently placed in an experimental 
setup may hinder attempts to study
macroscopic quantum coherence. Nevertheless, interesting results with
evidence for macroscopic quantum tunneling, energy level quantization and 
coherent dynamics between quantum levels were obtained with systems 
where the Josephson phase coordinate is trapped
in a metastable well (for an overview see
\cite{likharev83,clarke88,han96,silvestrini97,han02,martinis02}).
Also with systems where the energy scale for single-charge effects is 
higher than, or comparable to the energy scale for the Josephson effect, 
quantum coherent dynamics has recently been demonstrated \cite{nakamura99,saclay02}.
In Josephson junction loops, quantum superposition states of persistent currents
have been demonstrated spectroscopically
\cite{friedman2000,vanderwal2000}. However, time-resolved experiments that prove
quantum-coherent oscillations between macroscopically-distinct
persistent-currents states in the sense of 
Ref.\ \cite{leggett85} have not been reported yet.
   
Whether such experiments can be realized at all has been intensively
discussed in the literature \cite{tesche90}, without consensus being
reached. However, a detailed analysis with estimates based on measurement
techniques that can be realized in experiments {\it today}, has been
discussed very little. The quantum coherent dynamics observed with
the other Josephson junction systems (such as Cooper pair boxes
\cite{nakamura99,saclay02} or single junctions \cite{han02,martinis02}), 
indicates that it might be possible to obtain similar experimental 
results with Josephson persistent-current loops.
Efforts in this direction were stimulated by the
prospect that it might be possible to realize a quantum computer with
superconducting Josephson devices 
\cite{bocko97,ioffe99,makhlin99,mooij99,orlando99,makhlin2000}.
An important advantage of a Josephson quantum computer would be that, if
accurate quantum coherent control of elementary units would be possible,
it would be a system that can be extended to one containing a very large
number of quantum bits (qubits). The large size of the qubits allows
for individual (local) control and readout of the qubits and
qubit-qubit couplings.

In this article we analyze the feasibility of demonstrating
quantum coherent dynamics of Josephson persistent currents with
experimental techniques for manipulating and reading qubit states
that can be realized in the laboratories {\it today}
(i.~e.~assuming the available techniques for device fabrication,
cryogenics, microwave applications and electronic filtering). 
Such mesoscopic solid-state experiments suffer
from the fundamental difficulty that one cannot avoid that an electronic 
measuring device is permanently coupled to the {\it single} quantum system that is
studied \cite{manyloop}. We will not consider future measurement techniques 
which may couple less directly to the qubit. 
A meter must be present in any useful experiment, and, unlike
experiments with for instance photons, this means that a measuring device
must be permanently located very close to the solid-state quantum bit
(e.~g.~fabricated on the same chip). With such a setup, there is obviously a
conflict between an efficient measurement scheme with a strong measurement,
and long decoherence times in the quantum system that is studied. For
successful experiments in this direction, a detailed understanding of the
measurement scheme is therefore needed such that the decoherence that is
induced by the setup itself can be reduced to an acceptable level.

Obviously, there exist many other sources of decoherence for Josephson
qubits that one should worry about as well. The critical current of
the junctions may show telegraph noise \cite{harlingen1987},
which would give rise to decoherence similar to what is described in 
Ref.\  \cite{paladino2002}. 
Moreover, it has been
stressed that a very high number of spin degrees of freedom is usually
present in the solid state environment that may decohere
Josephson qubits (see the work by Prokof'ev and Stamp 
\cite{prokofev2000a,prokofev2000b} on the spin-bath, and Ref.\ \cite{tian2000} 
for estimates for persistent-current qubits). Another 
example is decoherence from quasiparticles that effectively shunt the
junction \cite{AES}. These effects themselves are very interesting for further 
study. However, a study of for example the dephasing due to spin
impurities remains impossible as long a reliable and well-understood
measurement scheme for the
loop's quantum dynamics is not available. Therefore, we will concentrate
here on dephasing and mixing due to the experimental wiring and the
measurement scheme itself.

\begin{figure}[htb]
\includegraphics[width=0.8\columnwidth]{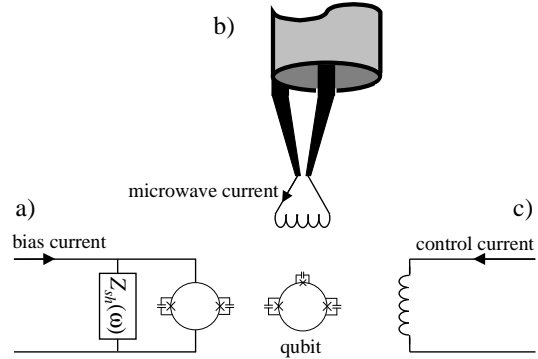}
\caption{Experimental setup for measurements on a Josephson
persistent-current qubit. The qubit (center) is a superconducting
loop that contains three Josephson junctions. It is inductively
coupled to a DC-SQUID (a), and superconducting control lines for
applying magnetic fields at microwave frequencies (b) and static
magnetic fields (c). The DC-SQUID is realized with an on-chip
shunt circuit with impedance $Z_{sh}$. The circuits a)-c) are
connected to filtering and electronics (not drawn).}
\label{figexp}
\end{figure}

Our analysis mainly focuses on experiments with the three-junction
persistent-current qubit proposed by Mooij {\it et al.},
\cite{mooij99,orlando99,vanderwal2000},
in a setup where they are measured by
underdamped DC-SQUID magnetometers (in this article we will reserve the
word SQUID for the measuring DC-SQUID (Fig.~\ref{figexp}a),
and not use it for the three-junction qubit (Fig.~\ref{figexp}, center)).
The decohering influence of the
inductively coupled DC-SQUID is analyzed as well as decoherence that results
from inductive coupling to on-chip control lines for applying microwave
signals and local magnetic fields. Model descriptions of the experimental
setup will be mapped on the spin-boson model, such that we can use
expressions for the relaxation and dephasing rates from the spin-boson
literature. The typical experimental situations will be described 
quite extensively to justify the models and the approximations
used. The results will be worked out quantitatively, and we will
evaluate whether we can realize mixing and dephasing rates that are
compatible with measurement schemes based on 
DC-SQUIDs. 
The design criteria developed in this work are more general and should 
also be of interest for experiments on loops with a single Josephson
junction \cite{friedman2000}, and quantum circuits where
the charge degree of freedom is measured, as Josephson charge quantum bits
\cite{makhlin99,makhlin2000}\ and quantum dots \cite{oosterkamp98}. 
In a more general
context the value of this work is that it presents in detail an
example of a measurement process on a single quantum system in which the
decoherence enhances with increasing measurement strength. The issues
discussed here are an example of experimental difficulties that will
unavoidably play a role in many realizations of quantum computers.

\subsection{Outline}

In section 2 we will summarize a theoretical description of the
Josephson persistent-current qubit, and the spin-boson theory that
will be applied in our analysis. Section 3 presents a description
of the measurement process with the DC-SQUIDs, and a typical
scheme for coupling the qubit to the on-chip control lines. In
section 4 we work out the qubit's relaxation and dephasing rate
that result from the coupling to a switching DC-SQUID. This is
worked out quite extensively and the definitions presented in this
section are also used in section 5. Two
measurement scenarios with different types of electromagnetic
shunt circuits for the DC-SQUID will be compared. A short analysis
of the decoherence due to the coupling to on-chip control lines is
presented in section 5. Section 6 presents a few control techniques 
that can improve decoherence rates.


\section{Qubit Hamiltonian and theory for relaxation and dephasing}

This work aims at calculating relaxation (mixing) rates and dephasing
(decoherence) rates for a Josephson persistent-current qubit which result
from its inductive coupling to the measurement setup. The measurement setup
is formed by a DC-SQUID and control lines, which are attached to leads and
coupled to filters and electronics (Fig.~\ref{figexp}). This setup will be
modeled as a macroscopic quantum two-level system (central spin) 
that is coupled to a linear electromagnetic impedance $Z_{t}(\omega )$, where 
$\omega$ the angular frequency.
The impedance $Z_{t}(\omega )$ forms an oscillator bath and
can be described by a set of $LC$ oscillators. 
This allows for mapping the problem on the spin-boson model: 
a central spin-$\frac12$ system that is
coupled to a bosonic bath \cite{leggett87,weiss99}. 
The parameters of the bath will be derived from the Johnson-Nyquist noise
from $Z_{t}(\omega )$.
In this section we will first
introduce the qubit Hamiltonian and physical properties of the qubit, and
then summarize the spin-boson expressions for relaxation and dephasing.

\subsection{Qubit properties and Hamiltonian}

The three-Josephson junction qubit \cite{mooij99,orlando99,vanderwal2000} is
a low-inductance superconducting loop which contains three Josephson tunnel
junctions (Fig.~\ref{figexp}). By applying an external flux $\Phi _{q}$ a
persistent supercurrent can be induced in the loop. For values where $\Phi
_{q}$ is close to a half-integer number of superconducting flux quantums
$\Phi _{0}$, two states with persistent currents of opposite sign are nearly
degenerate but separated by an energy barrier. We will assume here that the
system is operated near $\Phi _{q}=\frac12 \Phi _{0}$. Classically, the
persistent currents have here a magnitude $I_{p} $. Tunneling through the
barrier causes a weak coupling between the two states, and at low 
energies the loop can be
described by a Hamiltonian in the form of a two-level system 
\cite{mooij99,orlando99,vanderwal2000},
\begin{equation}
\hat{H}_{q}=\frac{\varepsilon }{2}\hat{\sigma}_{z}+\frac{\Delta }{2}\hat{\sigma}_{x},
\label{2lsHam}
\end{equation}
where $\hat{\sigma}_{z}$ and $\hat{\sigma}_{x}$ are Pauli spin operators. The two
eigen vectors of $\hat{\sigma}_{z}$ correspond to states that have a left or a
right circulating current and will be denoted as $|L\rangle $ and $|R\rangle
$. The energy bias $\varepsilon = 2I_{p}(\Phi _{q}-\frac12 \Phi _{0})$ is
controlled by the externally applied field $\Phi _{q}$. We follow
\cite{grifoni99} and define $\Delta $ as the tunnel splitting at $\Phi _{q}=
\frac12 \Phi _{0}$, such that $\Delta =2W$ with $W$ the tunnel coupling between
the persistent-current states. This system has two energy eigen values $\pm
\frac{1}{2}\sqrt{\Delta ^{2}+\varepsilon ^{2}}$, such that the level
separation $\nu $ gives
\begin{equation}
\nu =\sqrt{\Delta ^{2}+\varepsilon ^{2}}.
\end{equation}
In general $\Delta$ is a function of $\varepsilon$. 
However, it varies on the scale of the single junction plasma frequency, which is
much above the typical energy range at which the qubit is operated, such 
that we can assume $\Delta$ to be constant for the purpose of this paper. 

In the experiments $\Phi _{q}$ can be controlled by applying a magnetic
field with a large superconducting coil at a large distance from the qubit,
but for local control one can apply currents to superconducting control
lines, fabricated on-chip in the direct vicinity of the qubit. The qubit's
quantum dynamics will be controlled with resonant microwave pulses
(i.~e.~by Rabi oscillations). The proposed operation point is at
$\varepsilon \approx 5 \Delta$, which was analyzed to be a good 
trade-off between a system with significant tunneling, and 
a system with $\sigma_{z}$-like eigen states that can be
used for qubit-qubit couplings and measuring qubit states 
\cite{mooij99,orlando99}.
For optimal microwave control the qubit will be placed in a small off-resonant 
cavity, and the
microwave signals will be applied through on-chip superconducting
control lines (i.~e.~the magnetic component of
the fields from microwave currents will be
used). 
The qubit has a magnetic dipole moment as a result of
the clockwise or counter-clockwise persistent current
The corresponding flux in the loop is much smaller than the 
applied flux $\Phi_{q}$, but large enough to be detected
with a SQUID. 
This will be used for measuring the
qubit states. For our two-level system Eq.~\ref{2lsHam}, this means that 
both manipulation and readout couple to $\hat{\sigma}_{z}$. Consequently,
the noise produced by the necessary circuitry will couple in as
flux noise and hence 
couple to $\hat{\sigma}_{z}$, giving $\epsilon$ a small, stochastically time-dependent
part $\delta\epsilon (t)$. 
Our system also has electric dipole moments, represented by
$\hat{\sigma}_{x}$. These couple much less to the circuitry and
will hence not be discussed here.

\subsection{Spin-boson theory for relaxation and dephasing}

For defining the relaxation and dephasing rates, the state of the qubit is
described with a reduced density matrix $\overline{\rho }$, in the basis
which is spanned by the eigen vectors of $\hat{\sigma}_{z}$ in (\ref{2lsHam}),
i.e.\ by the semiclassical states with well-defined left ($L$) or right ($R$)
circulating current
\begin{equation}
\overline{\rho }=\left(
\begin{array}{cc}
\rho _{L,L} & \rho _{R,L} \\
\rho _{L,R} & \rho _{R,R}
\end{array}
\right) .
\end{equation}
We will concentrate our discussion on the undriven case. 
The qubit dynamics consists of quantum-coherent oscillations, which
decay on a time-scale $\tau_{\rm \phi}=\Gamma^{-1}_{\rm\phi}$, 
the dephasing time. This dephasing is superimposed
on an energy relaxation mechanism on a larger timescale 
$\tau_{\rm r}=\Gamma_{\rm r}^{-1}$, the relaxation time. 
This combined decoherence process
brings the system into an incoherent thermal mixture of its energy 
eigen states. 
Expressed in the basis of these eigen states, 
the off-diagonal terms (coherences)
of the density matrix $\overline{\rho}$ go to zero on the time scale of  
$\tau_{\rm\phi}$, whereas
the diagonal terms (populations) decay in $\tau_r$ to the 
Boltzmann factors.
For estimating $\Gamma _{r}$ and $\Gamma _{\phi }$ we will work from
the systematic weak-damping approximation (SWDA) developed
by Grifoni {\it et al.} \cite{grifoni99}, which covers recent theoretical progress
for the spin-boson theory.
Grifoni {\it et al.} calculated
expressions for $\Gamma _{r}$ and $\Gamma _{\phi }$ for a
spin-boson system in which the coupling to the environment is
dominated by bilinear coupling terms between $\hat{\sigma}_{z}$ and the
bath coordinates. This is a good approximation for a quantum
two-level system that is only weakly damped by the environment.

In our case the bath is formed by the impedance $Z_{t}(\omega )$, and
can be described by a set of $LC$ oscillators with flux
coordinates $\hat{\Phi}_{i}$, conjugate charge coordinates $\hat{Q}_{i}$, and
Hamiltonian 
\begin{equation}
\hat{H}_{bath}=\sum \nolimits_{i} 
\left( \hat{\Phi}_{i}^{2}/2L_{i}+\hat{Q}_{i}^{2}/2C_{i}\right) . 
\end{equation}
The flux produced by
the qubit will shift the flux $\hat{\Phi}_{i}$ in each $LC$ oscillator.
The coupling Hamiltonian is 
\begin{equation}
\hat{H}_{q-bath}=\frac{\hat{\sigma}_{z}}{2} \sum\nolimits_{i}c_{i}\hat{\Phi}_{i},
\end{equation}
where $c_{i}$ is the coupling
strength to the $i$-th oscillator. In this model the influence of
the oscillator bath on the qubit can be captured in the
environmental spectral density function
\begin{equation}
J(\omega )=\frac{\pi }{2 \hbar} \sum\nolimits_{i}
\left( c_{i}^{2}/C_{i}\omega _{i}\right)
\delta (\omega - \omega _{i}),
\label{Jw}
\end{equation}
where $\omega _{i}$ the resonance frequency of the $i$-th
oscillator. The dense spectrum of the degrees of freedom in the
electromagnetic environment allows for treating $J(\omega )$ as a
continuous function.

From now on, we focus on the low-damping limit, $J(\omega)\ll\omega$.
Thus, the energy-eigenstates of the qubit Hamiltonian, Eq.\ 
\ref{2lsHam}, are 
the appropriate starting point of our discussion.
In this case,
the relaxation rate $\Gamma _{r}$ (and relaxation time $\tau _{r}$) are
determined by the environmental spectral density $J(\omega )$ at the frequency
of the level separation $\nu $ of the qubit
\begin{equation}
\Gamma _{r}=\tau _{r}^{-1}=\frac{1}{2}\left( \frac{\Delta }{\nu }\right)
^{2}J(\nu /\hbar )\coth \left( \frac{\nu }{2k_{B}T}\right) ,  
\label{Gmix}
\end{equation}
where $T$ is the temperature of the bath. The dephasing rate $\Gamma _{\phi
} $ (and dephasing time $\tau _{\phi }$) is 
\begin{equation}
\Gamma _{\phi }=\tau _{\phi }^{-1}=\frac{\Gamma _{r}}{2}+
\left( \frac{\varepsilon }{\nu }\right) ^{2}\,
\alpha \,2\pi \frac{k_{B}T}{\hbar }.
\label{Gphase}
\end{equation}
These expressions have been derived in the context of NMR \cite{abragam61}
using a Markov approximation and recently been confirmed by a full 
path-inegral analysis \cite{grifoni99}.

The second term only contributes for an environment which is 
Ohmic at low frequencies (i.~e.~for $J(\omega )\propto \omega $). 
Here $\alpha $ is a dimensionless dissipation
parameter. It is determined by the slope of $J(\omega )$ at low frequencies
\begin{equation}
\alpha =  \lim_{\omega\rightarrow 0} \frac{J(\omega)}{2\pi\omega},
\label{alpha}
\end{equation}
which, if $J(\omega )$ is a sufficiently smooth function of $\omega$ 
can usually be taken as $\alpha = \frac{1}{2\pi}
\frac{\partial J(\omega )}{\partial \omega }$ at  $\omega \approx 0$. 
These results can be intuitively interpreted: The system can
relax by dissipating all its energy $\nu$ into an environmental boson. 
Due to the weakness of the coupling, there are no multi-boson processes. 
The relaxation also dephases the state. Moreover, dephasing can occur
due to the coupling to low-frequency modes which do not change the energy
of the system. These expressions for relaxation and dephasing have also been
found by studying the Hamiltonian of our qubit coupled to a damped
oscillator, using a Markovian master equation approach by Tian
{\it et al.} \cite{tian2001} (based on work by Garg {\it et al.}
\cite{garg85}).

The expressions (\ref{Gmix}) and (\ref{Gphase}) have prefactors $\left(
\frac{\Delta }{\nu }\right) ^{2}$ and $\left( \frac{\varepsilon }{\nu }
\right) ^{2}$ that depend on the tunnel splitting $\Delta $ and the energy
bias $\varepsilon $. These factors correspond to the angles
between noise and eigen states usually introduced in NMR \cite{abragam61} and
account for the effect that the qubit's
magnetic dipole radiation is strongest where the flux in the qubit
$\Phi_{q}= \frac12 \Phi _{0}$ (i.~e.~$\left( \frac{\Delta }{\nu }\right) $
maximal), and that the level separation $\nu $ is insensitive to flux noise
at this point (i.~e.~$\frac{\partial \nu }{\partial \varepsilon }=\left(
\frac{\varepsilon } {\nu }\right) \approx 0$). One should know and control
$J(\omega )$ at the frequency $\nu /\hbar $ for controlling the relaxation,
and at low frequencies for controlling the dephasing. In
this article we will calculate the noise properties of a few typical
experimental environments, and calculate how the noise couples to the qubit.
This can be used to define $J(\omega )$ for our specific environments.


\section{Measurement setup}

This section describes a typical experimental setup for measurements on Josephson
persistent-current qubits. Only the parts that are most strongly coupled to
the qubit will be worked out (Fig.~\ref{figexp}). The first part
describes a DC-SQUID magnetometer that is used by measuring its
switching current, the second part addresses the use of on-chip
superconducting lines for applying magnetic fields to the qubit.

\subsection{Switching DC-SQUID}

\begin{figure}[htb]
\begin{minipage}{0.9\columnwidth}
\includegraphics[width=0.8\columnwidth]{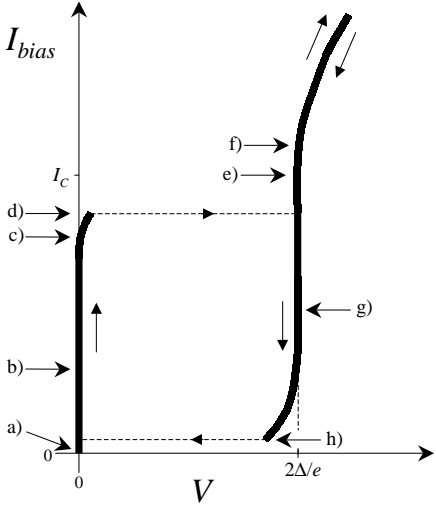}
\caption{Sketch of a typical hysteretic current-voltage
characteristic ($IV$) for a current-biased Josephson junction or
small DC-SQUID. The $IV$ is hysteretic; arrows indicate which of
the two branches is followed at an increase or decrease of the
bias current. When the bias current $I_{bias}$ is ramped up from
zero (a), the voltage $V$ first remains zero. The circuit is here
on the supercurrent branch of the $IV$ (b). When $I_{bias}$
approaches the critical current $I_{C}$, a slow diffusive motion
of the phase $\protect\varphi_{ext}$ leads to a very small voltage
across the system (c). At slightly higher current (d), but always
below $I_{C}$ (e), the system switches to a running mode for
$\protect\varphi_{ext}$, and the voltage jumps to a value set by
quasiparticle tunneling over the superconducting gap, $V = 2\Delta
/ e$ (this current level (d) is the switching current $I_{SW}$).
At further increase of the current (f) the $IV$ approaches an
Ohmic branch, where transport is dominated by quasiparticle
tunneling through the normal tunnel resistance of the junctions.
When lowering the bias current the system follows the running mode
(g) down to a low bias current where it retraps on the
supercurrent branch (at the level $I_{retrap}$, indicated by (h)).
See also the corresponding washboard potential model, in
Fig.~\ref{figwash}.} \label{figiv}
\end{minipage}
\end{figure}

SQUIDs are the most sensitive magnetometers, and they can be operated at
very low power consumption \cite{tinkham96}. We will consider here the use
of a DC-SQUID with a hysteretic current-voltage characteristic ($IV$),
and unshunted junctions that are extremely underdamped.
It is used by ramping a current
through it and recording the switching current: the bias current at which it
switches from the supercurrent branch to a nonzero voltage in its $IV$
(Fig.~\ref{figiv}). The switching current is a measure for the magnetic flux in
the loop of the SQUID. An important advantage of this scheme is that the
SQUID produces before readout very little noise. As long as the SQUID is on
the supercurrent branch, it does not produce any shot noise or Josephson
oscillations. If the external noise and interference can be suppressed by
filtering, there is only Johnson-Nyquist noise from the low-temperature
leads and filtering that the SQUID is connected to. At low frequencies this
residual noise has little power since the device is superconducting.
Moreover, we will show in section 4 that at low bias currents the effective
coupling between this meter and the quantum system is very weak. In
comparison, damped non-hysteretic SQUIDs have the problem that the shunt
resistors at the junctions also provide a damping mechanism for the qubit.
In a hysteretic SQUID there is more freedom to engineer the effective
impedance seen by the qubit, and it also has the advantage that the voltage
jump at the switching current is much larger \cite{joyez99}. 
Recently, a similar scheme with a superconducting single-charge
device, that can be operated as a switching electrometer  
has been reported \cite{cottet02,saclay02}.
Voltage biased single-electron transistors for quantum measurements have been 
analyzed in Refs.~\cite{schnirman98,devoret2000,zorin2002,korotkov2001}.

\begin{figure}[htb]
\begin{minipage}{0.9\columnwidth}
\includegraphics[width=0.8\columnwidth]{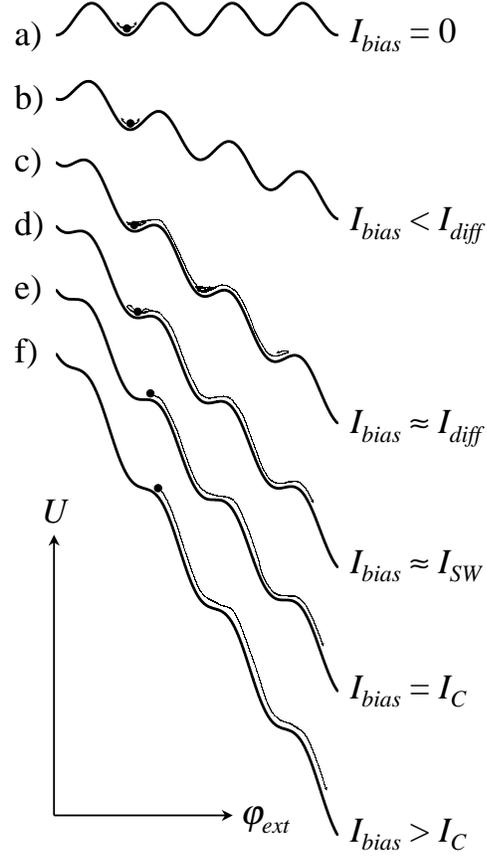}
\caption{The dynamics of a current-biased DC-SQUID, modeled as a
particle with coordinate $\protect\varphi_{ext}$ in a
one-dimensional tilted washboard potential $U$. The labeling a)-f)
corresponds to that of Fig.~\ref{figiv}. At zero (a) and small
bias currents (b), the particle is trapped in a well of the
washboard. Apart from the small plasma oscillations at the bottom
of the well, the particle's coordinate $\protect\varphi_{ext}$ is
fixed. When increasing the slope of the washboard, the particle
will start to have a slow, on average downwards, diffusive
dynamics, with rare excursions to one of the neighboring wells
(c). At the switching current $I_{SW}$ there is a high probability
that the trapped particle will escape to a running mode (d), with
effectively zero probability for retrapping. Here the loss of
potential energy exceeds the dissipation when the particle moves
one period down the washboard, and the particle builds up a high
kinetic energy. Due to thermal fluctuations, external noise, and
in certain cases quantum fluctuations, this occurs below the
critical current $I_{C}$: the slope where all local minimums in
the washboard potential disappear (e). At currents higher than
this slope (f), the particle will always be in a running mode. The
retrapping process when lowering the bias current follows similar
dynamics.} \label{figwash}
\end{minipage}
\end{figure}

For qualitative insight in the measurement process we will present here a
simplified description of the SQUIDs noise and dynamics (valid for a
DC-SQUID with symmetric junctions and a loop with negligible self
inductance). In section 4 it will be worked out in more detail. The
supercurrent through the SQUID with a flux $\Phi $ in its loop is
\begin{equation}
I_{sq}=2I_{co}\cos f\,\,\sin \varphi _{ext},  \label{IsqSym}
\end{equation}
where $f=\pi \Phi /\Phi _{0}$, $I_{co}$ the critical current of the
junctions, and $\varphi _{ext}$ a Josephson phase coordinate.
$I_{sq}$ will be distinguished from the applied bias current $I_{bias}$,
as part of the bias current may go into circuitry shunting the SQUID.
Insight in the SQUID's response to a bias
current is achieved by recognizing that (\ref
{IsqSym}) gives steady state solutions ($\partial U/\partial \varphi
_{ext}=0 $) for a particle with coordinate $\varphi _{ext}$, trapped on a
tilted washboard potential (Fig.~\ref{figwash})
\begin{equation}
U=-\frac{\hbar }{2e}\left( 2I_{co}\cos f\,\,\cos \varphi
_{ext}+I_{sq}\varphi _{ext}\right) .  \label{Uwashboard}
\end{equation}
In this picture, the average slope of the potential is proportional to the
bias current, and the supercurrent branch of the SQUID's $IV$ corresponds to
the particle being trapped in a well. The Josephson voltage across the SQUID
$V=\frac{\hbar }{2e}\frac{d\varphi _{ext}}{dt}$ is nonzero for the particle
in a running mode. In absence of noise and fluctuations, the SQUID will
switch to the running mode at the critical current $I_{C}$
\begin{equation}
I_{C}=2I_{co}\left| \cos f\right| .
\label{Ic}
\end{equation}
A DC-SQUID can thus be regarded as a single Josephson junction with a
flux-tunable critical current. In practice, noise and fluctuations of
$\varphi _{ext}$ will cause the SQUID to switch before the bias current
reaches $I_{C}$. This current level will be denoted as the switching current
$I_{SW}$ to distinguish it from $I_{C}$. It is a stochastic variable, but
averaging over repeated recordings of $I_{SW}$ allows for determining $f$
with great accuracy. This naive description can be used to illustrate three
important properties of the measurement process with the SQUID.

In the experiment, the electronics for recording the SQUID's $IV$ obtains
information about $f$ when the SQUID switches. However, rewriting (\ref
{IsqSym}) as
\begin{equation}
\varphi _{ext}=\sin ^{-1}\left( \frac{I_{sq}}{2I_{co}\cos f}\right)
\label{phiextcoor}
\end{equation}
shows that the SQUID's coordinate $\varphi _{ext}$ is already correlated 
(i.~e.~entangled) with the flux $f$ at current values below $I_{SW}$. Small voltage
fluctuations that result from small plasma oscillations and translations of
$\varphi _{ext}$ will cause dissipation in the electromagnetic environment of
the SQUID, which damps the dynamics of $\varphi _{ext}$. This means, that in
a quantum mechanical sense, the position of $\varphi _{ext}$, and thereby $f$,
is measured by the degrees of freedom that form the electromagnetic
impedance that is shunting the SQUID (i.~e.~the leads and filtering between
the SQUID and the readout electronics), and that the measurement may in fact
take place before it is recorded by a switching event.

Secondly, (\ref{phiextcoor}) shows that the SQUID's coordinate $\varphi
_{ext}$ is independent of the flux in the loop ($\partial \varphi
_{ext}/\partial f=0$) for $I_{sq}=0$. Therefore, in absence of fluctuations
of $\varphi _{ext}$ and current noise, the meter is at zero current
effectively ``off''. In practice this can not be perfectly realized, but it
illustrates that the decoherence from the SQUID may be reduced by a large
extent at low bias currents.

Thirdly, for bias currents well below $I_{C}$, the coordinate $\varphi
_{ext} $ is trapped in a potential that is for small oscillations close to
harmonic. The SQUID can in this case be regarded as an inductance
\begin{equation}
L_{J}=\frac{\hbar }{2e}\frac{1}{\sqrt{4I_{co}^{2}\cos ^{2}f-I_{sq}^{2}}}
\label{LjEarly}
\end{equation}
(see also (\ref{Lj}) below). The noise from the SQUID can here be described
by the Johnson-Nyquist noise from the SQUID's Josephson inductance (\ref
{LjEarly})\ in parallel with the SQUID's environmental impedance (Fig.~\ref
{figwash}a, b). For high bias currents very close to $I_{C}$, the spectrum
will have more power and calculating the noise properties will
be more complicated. Here non-harmonic
terms in the trapping potential become important, and there maybe additional
noise from a diffusive motion of $\varphi _{ext}$ to neighboring wells
(Fig.~\ref{figwash}c). For hysteretic SQUIDs this regime with
diffusive motion of $\varphi _{ext}$ and switching currents very close to
$I_{C}$ will only occur in SQUIDs with a very specific electromagnetic shunt
\cite{vion96,joyez99}. In many realizations of hysteretic DC-SQUIDs
$\varphi_{ext}$ will escape to a running mode without retrapping
in lower wells (Fig.~\ref{figwash}d), and $I_{SW}$ can be much lower than $I_{C}$.
In this case the approximation using (\ref {LjEarly})
should be valid for description of the noise before a switching
event.

The statistics of $I_{SW}$ readouts depend strongly on the damping of the
dynamics of $\varphi _{ext}$ by the impedance that is shunting the SQUID.
Experimental control over the damping, requires the fabrication of a shunt
circuit in the direct vicinity of the SQUID, such that its impedance is well
defined up to the frequency of the SQUID's plasma oscillations (microwave
frequencies). The shunt circuit is therefore preferably realized on-chip
($Z_{sh}$ in Fig.~\ref{figexp}a). The escape from the well may be thermally
activated, but for underdamped systems with low-capacitance junctions
quantum tunneling through the barrier can dominate the escape rate
at low temperatures. The influence of the damping circuitry on the
$I_{SW}$ statistics \cite{clarke88,vion96,joyez99} is now well understood.
A SQUID with very
underdamped dynamics usually has $I_{SW}$ values much below $I_{C}$, and
histograms of a set of $I_{SW}$ recordings will be very wide. This means
that one needs to average over many repeated measurements to achieve the
required resolution in readout. Thereby, averaging also needs to take place
over many repeated experiments on the qubit, such that only a time-ensemble
average can be measured. 
With a shunt that provides high damping at the
plasma frequency very narrow switching current histograms can be realized
\cite{vion96,joyez99,saclay02,orlando2002,tanaka2002}, that 
in principle allow for single-shot readout in qubit
experiments. While in such a scheme the SQUID's noise will also be
enhanced, it is possible to engineer (for realistic fabrication
parameters) a shunt impedance that is at the same
time compatible with coherent dynamics of the qubit and single-shot
readout \cite{cottet02}. The engineering of single-shot readout will not
be addressed in detail in this paper.

The main disadvantage of the switching SQUID is that it is not very
efficient. During each cycle through the hysteretic $IV$ it is only
measuring for a short time. Moreover, the $IV$ is very nonlinear, such that
the repetition frequency must be an order lower than the bandwidth of the
filters. The filtering that is required for realizing low effective
temperatures and the SQUID's shunt circuit have typically a bandwidth well
below 1GHz, and the accurate readout electronics set a similar limit to the
bandwidth. In practice this limits the repetition frequency to values in the
range of 10 kHz \cite{vanderwal2000,wernsdorfer2000}\ to 1 MHz
\cite{silvestrini97,cottet02}.
More efficient readout may be realized with AC readout techniques 
(see e.\ g.\ Ref.\  \cite{muck2001}).

The slow operation of the switching DC-SQUID sets requirements for the
mixing rate $\Gamma _{r}$ of the qubit. It needs be longer than the time
required to perform a switching current measurement, which requires a time
in the range 1 ${\rm \mu s}$ to 100 ${\rm \mu s}$. One could go to shorter
times by setting the SQUID ready at a high bias current when an experiments
on the qubit is started, but it is also needed to have the mixing time
longer than the time it takes to ramp the bias current through the range of
the switching current histogram. At the same time we should realize that the
quantum system is prepared by waiting for it to relax to the ground state,
so relaxation times very much longer than 100 ${\rm \mu s}$ will prohibit a
high repetition frequency. A high repetition frequency is needed if the
signal can only be build up by averaging over many switching events.

The experiments aim at working with many coherent Rabi oscillations with a
period of about 10 ns \cite{mooij99}. We therefore aim at engineering SQUIDs
that cause a dephasing time that is much longer than 10 ns. The dephasing
and relaxation times turn out to be shortest at high bias currents through
the SQUID. Unless mentioned otherwise, we will make in this article worst
case estimates for the dephasing and relaxation times using bias current
values near the switching current.

\subsection{On-chip control lines}

An attractive feature of macroscopic qubits is that one can address
individual qubits with control signals from microfabricated lines (see also
Fig.~\ref{figexp}b,c). For persistent-current qubits, for example, a
supercurrent through a line that is mainly coupled to one specific qubit can
be used for tuning this qubit's energy bias $\varepsilon $. Also, it is
convenient to provide the microwave signals for control of the qubit's
quantum dynamics using local superconducting lines. If this is realized in a
microwave cavity with its first resonance well above the applied microwave
frequency, one can apply microwave bursts with fast switch times without
being hindered by high-$Q$ electromagnetic modes in the volume that is
formed by the cold metallic shielding that surrounds the sample.

Microwave signals can be applied using external microwave sources at
room temperature. Alternatively, on-chip oscillators for example based on
Josephson junction circuits \cite{jain84,crankshaw2001} can be applied.
High microwave currents in the control lines are achieved by shorting the
microwave coax or wave guide close to the qubit with an inductance that has
an impedance much lower than the source's output impedance (Fig.~\ref{figexp}b).
For external microwave sources, the typical level for the output
impedance will be that of the available coax technology, typically
50 $\Omega $. With on chip Josephson oscillators the typical output impedance is
one order lower. In both cases, it is in practice very tedious
to engineer these impedance levels and our analysis below will show that this
forms a constraint for qubit experiments: long decoherence times are
in conflict with the wish for local qubit control and low power levels of
the applied microwave signals.

If one uses external microwave sources at room temperature it is harder than
for the quasi DC signals to filter out the high temperature noise. Low
effective temperatures can be achieved by a combination of narrow-band
microwave filters and strong attenuators at low temperatures.


\section{Relaxation and dephasing from a switching DC-SQUID}

\subsection{Current-phase relations for the DC-SQUID}

The DC-SQUID has two phase degrees of freedom, the gauge-invariant phases
$\gamma _{r}$ and $\gamma _{l}$ of the junctions \cite{tinkham96}. 
They are related to the supercurrents through the left and the right junction,
\begin{equation}
\begin{array}{c}
I_{l}=(I_{co}+\frac{\Delta I_{co}}{2})\sin \gamma _{l},  \\
I_{r}=(I_{co}-\frac{\Delta I_{co}}{2})\sin \gamma _{r}.
\end{array}
\end{equation}
Here $I_{co}$ is the average of the critical current of the two junctions. A
small asymmetry in the junctions' critical currents is accounted for by
$\Delta I_{co}\ll I_{co}$ (typically a few percent). We will work here with
the sum and difference phase coordinates $\varphi _{int}$ and $\varphi
_{ext} $, which are related by a linear transformation
\begin{equation}
\begin{array}{c}
\varphi _{ext}=\frac{\gamma _{l}+\gamma _{r}}{2} \\
\varphi _{int}=\frac{\gamma _{l}-\gamma _{r}}{2}
\end{array}
\;\;\;\;\Leftrightarrow \;\;\;\;
\begin{array}{c}
\gamma _{l}=\varphi _{ext}+\varphi _{int} \\
\gamma _{r}=\varphi _{ext}-\varphi _{int}
\end{array}
\;\;\;\;.
\end{equation}
The new phase coordinates are related with the current passing through the
SQUID $I_{sq}$ and the circulating current in the SQUID $I_{cir}$
\begin{equation}
\begin{array}{c}
I_{sq}=I_{l}+I_{r} \\
I_{cir}=\frac{I_{l}-I_{r}}{2}
\end{array}
\;\;\;\;\Leftrightarrow \;\;\;\;
\begin{array}{c}
I_{l}=\frac{1}{2}I_{sq}+I_{cir} \\
I_{r}=\frac{1}{2}I_{sq}-I_{cir}
\end{array}
\;\;\;\;,
\end{equation}
yielding the following current-phase relation for $I_{sq}$ and $I_{cir}$
\begin{eqnarray}
I_{sq} &=&2I_{co}\cos \varphi _{int}\sin \varphi _{ext}+\;\;\Delta
I_{co}\sin \varphi _{int}\cos \varphi _{ext},
\label{sqcur}\\
I_{cir} &=&\,\,\,I_{co}\sin \varphi _{int}\cos \varphi _{ext}+ \frac12
\Delta I_{co}\cos \varphi _{int}\sin \varphi _{ext}.
\label{circur}
\end{eqnarray}
We will assume that the DC-SQUID has junctions with a critical current and capacitance that
are lower than that of the qubit junctions. In this case, the internal phase
$\varphi _{int}$
follows the flux adiabatically up to time scales much faster than $\frac{\hbar }{\nu }$.
We will therefore use 
\begin{equation}
\varphi _{int}=\pi \frac{\Phi }{\Phi _{0}}\stackrel{def}{=}f.
\label{fluxoidqsq}
\end{equation}
\subsection{Noise on the qubit from the DC-SQUID resulting in $J(\protect \omega )$}

The noise that is induced by the measuring SQUID results from
Johnson-Nyquist noise of the total impedance $Z_{t}(\omega)$ between the leads
that are attached to the SQUID. The impedance $Z_{t}(\omega)$ is formed the SQUID's
impedance in parallel with the impedance of the wiring and circuitry that
the SQUID is connected to (see the circuit models in Fig.~\ref{figsqcir}).
At bias currents well below the critical current $I_{C}$, the
phase dynamics can be linearized and the SQUID can
be modeled as an inductor $L_{J}$. The coupling of $\varphi _{ext}$
to the SQUID's inner degree of freedom $\varphi_{int}$ and thereby
to the qubit slightly alter the effective value for $L_{J}$,
but the correction it is so small that it can be neglected.
The Fourier-transformed power spectrum 
$\langle \delta V(t)\;\delta V(0)\rangle _{\omega }$ of
the Johnson-Nyquist voltage fluctuations $\delta V$
across the SQUID is \cite{weiss99,devoret97}
\begin{equation}
\langle \delta V\;\delta V\rangle _{\omega }=\hbar \omega {\rm Re}
\{Z_{t}(\omega )\}\coth \left( \frac{\hbar \omega }{2k_{B}T} \right).
\label{JNnoise}
\end{equation}
We will now calculate how this voltage noise leads to fluctuations $\delta
\varepsilon $ of the energy bias on the qubit. As a rule, the spectral density
$J(\omega )$ in (\ref{Jw}) can then be derived by dividing the expression for
$\langle \delta \varepsilon \;\delta \varepsilon \rangle _{\omega }$
by $\hbar ^{2} \coth \left( \frac{\hbar \omega }{2k_{B}T}\right)$.

The current-phase relations for $I_{sq}$ and $I_{cir}$ can be used
for expressing the current fluctuations.
The first term of (\ref{sqcur}) gives
\begin{equation}
\begin{array}{c}
\frac{dI_{sq}}{dt}=i\omega I_{sq}\approx 2I_{co}\cos f
\cos \bar{\varphi}_{ext}\frac{d\varphi _{ext}}{dt} \\
=2I_{co}\cos f\cos \bar{\varphi}_{ext}
\frac{2e}{\hbar }V,
\end{array}
\label{dcurdt}
\end{equation}
where we used $\bar{\varphi}_{ext}$ for the time average
of $\varphi _{ext}$.
With a similar expression for the second term of (\ref{sqcur}) the
current fluctuations in $I_{sq}$ are
\begin{equation}
\delta I_{sq}\approx \left( 2I_{co}\cos f\cos \bar{\varphi}_{ext}-\Delta
I_{co}\sin f\sin \bar{\varphi}_{ext}\right) \;\delta \varphi _{ext}.
\label{sqcurnoise2}
\end{equation}
The SQUID is usually operated in regions where the average external flux in
its loop is between an integer and half-integer number of $\Phi _{0}$.
At these points $\left| \cos f\right| \approx \left| \sin f\right| $.
Therefore, the second term in (\ref{sqcurnoise2}) can be neglected unless
$\left| I_{co}\cos \bar{\varphi}_{ext}\right| \lessapprox
\left| \Delta I_{co}\sin \bar{\varphi}_{ext}\right| $.
That is, it can be neglected unless the bias current
is very high, for which $\sin \bar{\varphi}_{ext}$ approaches 1.
For most purposes we can thus use
\begin{equation}
\delta I_{sq}\approx 2I_{co}\cos f\cos \bar{\varphi}_{ext}\;\delta \varphi
_{ext}.
\label{sqcurnoise}
\end{equation}
This is also used to define $L_{J}$ by expressing
\begin{equation}
V=L_{J}\frac{dI_{sq}}{dt},
\end{equation}
such that with (\ref{dcurdt}), (\ref{fluxoidqsq}) and (\ref{sqcur})
$L_{J}$ should be defined
as
\begin{equation}
L_{J}=\frac{\hbar }{2e}\frac{1}{2I_{co}\cos f\cos \bar{\varphi}_{ext}}=
\frac{\hbar }{2e}\frac{1}{\sqrt{4I_{co}^{2}\cos ^{2}f-I_{sq}^{2}}}.
\label{Lj}
\end{equation}
For $I_{cir}$ we get a similar expression as (\ref{sqcurnoise2})
\begin{equation}
\delta I_{cir}\approx ( -I_{co}\sin f\sin \bar{\varphi}_{ext}+ \frac12
\Delta I_{co}\cos f\cos \bar{\varphi}_{ext} ) \;\delta \varphi _{ext}.
\label{circurnoise2}
\end{equation}
Using again that the SQUID is operated where
$\left| \cos f\right| \approx \left| \sin f\right| $
shows that the second term in (\ref{circurnoise2}) can be neglected unless
$\left| I_{co}\sin \bar{\varphi}_{ext}\right| \lessapprox \left| \Delta I_{co}\cos \bar{\varphi}_{ext}\right| $.
For $\delta I_{cir}$ the second term only plays a role at low bias currents in the
SQUID for which $\bar{\varphi}_{ext}\approx 0$, and for most
purposes we can use
\begin{equation}
\delta I_{cir}\approx -I_{co}\sin f\sin \bar{\varphi}_{ext}\;\delta \varphi
_{ext}.
\label{circurnoise}
\end{equation}
In the above we used $\bar{\varphi}_{ext}$ for the time average
of $\varphi _{ext}$, but at places where it is not confusing it
will be simply denoted as $\varphi _{ext}$.

Both noise in $I_{sq}$ and $I_{cir}$ can couple to the qubit,
but we will assume that the qubit is mainly sensitive to
noise in $I_{cir}$ (as in the experiments in \cite{vanderwal2000},
where the qubit was placed symmetrically inside the SQUID's loop)
and neglect an inductive coupling to noise in $I_{sq}$.
For a more general approach, coupling to noise in $I_{sq}$ can be treated
on a similar footing as noise in $I_{cir}$, but for all useful
sample geometries it should give a contribution to relaxation
and dephasing rates that is at most
on the same order as that of $I_{cir}$.

With
$i\omega \delta I_{cir}=-\frac{2e}{\hbar }
I_{co}\sin f\sin \varphi_{ext}\;\delta V$
follows for the fluctuations
$\delta I_{cir}$
\begin{equation}
\langle \delta I_{cir}\;\delta I_{cir}\rangle _{\omega }=
\left( \frac{2e}{\hbar }\right) ^{2}
\frac{1}{\omega ^{2}}I_{co}^{2}\sin ^{2}f\sin ^{2}\varphi_{ext}\;
\langle \delta V\;\delta V\rangle _{\omega } \;\;.
\end{equation}
The fluctuations in the imposed qubit flux are $\delta \Phi _{q}=M\delta I_{cir}$,
where $M$ the mutual inductance between the SQUID loop and the qubit loop.
This then yields the fluctuations in the energy bias with
$\delta \varepsilon = 2I_{p}\delta \Phi _{q}$, 
\begin{equation}
\langle \delta \varepsilon \;\delta \varepsilon \rangle _{\omega }=\left(
\frac{2e}{\hbar }\right) ^{2}\frac{4}{\omega ^{2}}M^{2}I_{p}^{2}I_{co}^{2}
\sin ^{2}f\sin ^{2}\varphi _{ext}\;\langle \delta V\;
\delta V\rangle_{\omega }
\end{equation}
where $I_{\rm p}$ is the amplitude of the circulating current in the qubit
in the semiclassical states.
Using (\ref{sqcur}) and (\ref{JNnoise}) and filling in
$\frac{h}{2e}=\Phi_{0}$ this can be written as
\begin{equation}
\begin{array}{c}
\langle \delta \varepsilon \;\delta \varepsilon \rangle _{\omega }=\hbar
\left( 2\pi \right) ^{2}\frac{1}{\omega }\left( \frac{MI_{p}}{\Phi _{0}}\right) ^{2}
I_{sq}^{2}\tan ^{2}f\; \times \\
{\rm Re} \{Z_{t}(\omega )\}\coth \left(
\frac{\hbar \omega }{2k_{B}T}\right)
\end{array}
\label{levelnoise}
\end{equation}

The fluctuations
$\langle \delta \varepsilon \;\delta \varepsilon \rangle _{\omega }$
are the result of the coupling to the oscillator bath, as in
(\ref{Jw}). This can be
used to define $J(\omega)$ for our specific environment,
\begin{equation}
J(\omega )=\frac{\left( 2\pi \right) ^{2}}{\hbar }\frac{1}{\omega }\left(
\frac{MI_{p}}{\Phi _{0}}\right) ^{2}I_{sq}^{2}\tan ^{2}f\; {\rm Re}
\{Z_{t}(\omega )\} .
\label{Jwsquid}
\end{equation}
These results show, that although the SQUID is permanently close
to the qubit, it may be effectively decoupled if there is no net 
bias current $I_{sq}$ flowing through the device. The physical reason for
this becomes apparent in Eqs. (\ref{sqcur}) and (\ref{circur}): The SQUID remains mirror symmetric in 
that case and consequently the fluctuations of the bias current are diverted
symmetrically around the arms of the SQUID and do not produce flux
noise \cite{insquid}.

\subsection{Relaxation times}

With (\ref{Gmix}) and (\ref{Jwsquid}) follows the SQUID's contribution
to the relaxation rate.
It is here expressed as a function of the resonance frequency
$\omega_{res}=\nu /\hbar $ at which the qubit is operated,
\begin{equation}
\begin{array}{c}
\Gamma _{r}=\left( \frac{\Delta /\hbar }{\omega_{res}}\right) ^{2}
\frac{\left( 2\pi \right) ^{2}}{2\hbar }\frac{1}{\omega_{res}}
\left( \frac{MI_{p}}{\Phi _{0}}\right) ^{2}I_{sq}^{2}\tan ^{2}f\; \times \\
{\rm Re} \{Z_{t}(\omega_{res})\}
\coth \left( \frac{\hbar \omega_{res}}{2k_{B}T}\right) .
\end{array}
\label{Grsquid}
\end{equation}
In this formula one can recognize a dimensionless factor
$\left( \frac{MI_{p}}{\Phi _{0}}\right) ^{2}$
which is a scale for how strongly the qubit is
coupled to the measuring SQUID. A dissipation factor in the form $I^{2}R$
can be recognized in $I_{sq}^{2}\tan ^{2}f\; {\rm Re} \{Z_{t}(\omega )\}$.
The dissipation scales with the absolute value of the current fluctuations,
so with $I_{sq}$, and the expression is independent of the critical current of
the SQUID junctions $I_{co}$ (unless ${\rm Re} \{Z_{t}(\omega )\}$ depends
on $I_{co}$).
A weak measurement scheme in which the inductive coupling
to a DC-SQUID $(MI_{p}/\Phi _{0})^{2} << 1 $ can
yield relaxation rates that are very low when compared to
a scheme in which leads are directly attached to the
loop \cite{threejunctionsrefs}. A measurement of such a
scheme's switching current could also be used
for probing the qubit, but the influence of the voltage
noise would be dramatically worse.

With the result (\ref{Grsquid}) the relaxation
rate for typical sample parameters will be calculated.
Sample parameters similar to our recent
experiment \cite{vanderwal2000} are
$\omega_{res}=10\;{\rm GHz}$,
$\Delta =2\;{\rm GHz}$,
$\frac{MI_{p}}{\Phi _{0}}=0.002$.
It is assumed that a SQUID with
$2I_{co}=200\;{\rm nA}$ is operated at
$f=0.75\;\pi $ and biased near
the switching current, at $I_{sq}=120\;{\rm nA}$.
For $T=30\;{\rm mK}$ the relaxation rate per Ohm environmental
impedance is then
\begin{equation}
\tau _{r}=\Gamma _{r}^{-1}\approx \frac{150\;{\rm \mu s\,\Omega }}{{\rm Re}
\{Z_{t}(\omega_{res})\}}.  \label{mixest}
\end{equation}

\subsection{Engineering ${\rm Re} \{Z_{t}(\protect\omega )\}$ for slow 
relaxation}

\begin{figure}[htb]
\begin{minipage}{0.9\columnwidth}
\vspace{5mm}
\includegraphics[width=0.8\columnwidth]{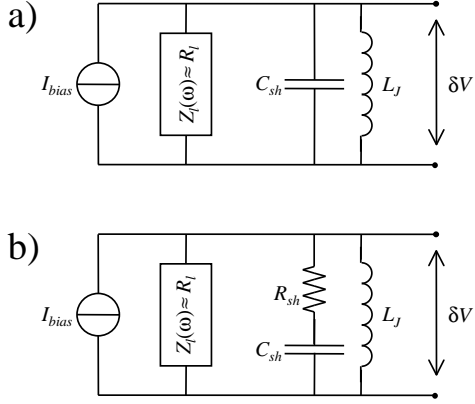}
\caption{Circuit models for the $C$-shunted DC-SQUID (a) and the
$RC$-shunted DC-SQUID (b). The SQUID is modeled as an inductance
$L_J$. A shunt circuit, the superconducting capacitor $C_{sh}$ or
the $R_{sh}$-$C_{sh}$ series, is fabricated on chip very close to
the SQUID. The noise that couples to the qubit results from
Johnson-Nyquist voltage noise $\protect \delta V$ from the
circuit's total impedance $Z_t$. $Z_t$ is formed by a parallel
combination of the impedances of the leads $Z_l$, the shunt and
the SQUID, such that $Z_t=(1/Z_l+1/(R_{sh}+1/i \protect\omega
C_{sh}) + 1/i \protect\omega L_j)^{-1}$, with $R_{sh} = 0$ for
(a).} \label{figsqcir}
\end{minipage}
\end{figure}

In practice the SQUID's resolution is improved by building an on-chip
electromagnetic environment. We will consider here
a large superconducting capacitive shunt (Fig.~\ref{figsqcir}a,
as in our recent experiment \cite{vanderwal2000}).
This scheme will be denoted as the $C$-shunted SQUID. As an alternative we
will consider a shunt that is a series combination of a
large capacitor and a resistor (Fig.~\ref{figsqcir}b).
This will be denoted as the $RC$-shunted SQUID.
The $C$ shunt only makes the effective mass of
the SQUID's external phase $\varphi _{ext}$ very heavy. The $RC$ shunt also
adds damping at the plasma frequency of the SQUID, which is needed
for realizing a high resolution of the SQUID readout
(i.~e.~for narrow switching-current histograms) \cite{joyez99}.
The total impedance $Z_{t}(\omega)$ of the two measurement circuits are modeled
as in Fig.~\ref{figsqcir}.
We assume a perfect current source $I_{bias}$ that ramps the
current through the SQUID. The fact that the current source is
non-ideal, and that
the wiring to the SQUID chip has an impedance is all modeled by the
impedance $Z_{l}$. The wiring can be engineered such that for a very wide
frequency range the impedance $Z_{l}$ is on the order of the vacuum
impedance, and can be modeled by its real part $R_{l}$. It typically has
a value of $100\;\Omega $. On chip, the impedance is formed by 
the Josephson inductance $L_{J}$
in parallel with the shunt circuit ($C_{sh}$, or the series combination
of $R_{sh}$ and $C_{sh}$).
We thus assume that the total impedance $Z_{t}(\omega)$ can be
described as
\begin{equation}
Z_{t}(\omega)=\left( \frac{1}{i\omega L_{J}}+
\frac{1}{\frac{1}{i\omega C_{sh}}+R_{sh}}+
\frac{1}{R_{l}}\right) ^{-1},
\end{equation}
where $R_{sh}$ should be taken zero for the $C$-shunt scenario.

The circuits in Fig.~\ref{figsqcir} are damped $LC$ resonators. It is clear
from (\ref{Gmix}) and (\ref{Jwsquid}) that one should keep the
$LC$-resonance frequency $\omega _{LC}=1/\sqrt{L_{J}C_{sh}}$, where ${\rm Re}
\{Z_{t}(\omega )\}$ has a maximum, away from the qubit's resonance $\omega
_{res}=\nu /\hbar $. For practical values this requires
$\omega _{LC}\ll \omega_{res}$ for Aluminum technology (with Niobium-based 
technology, the regime
$\omega _{LC}\gg \omega_{res}$ may be realized \cite{orlando2002}). 
This then gives the circuits
a ${\rm Re} \{Z_{t}(\omega )\}$ and $J(\omega )$ as plotted in
Fig.~\ref{figsqrez}.
For the circuit with the $C$ shunt
\begin{equation}
{\rm Re} \{Z_{t}(\omega)\}\approx
\begin{array}{cc}
\frac{\omega ^{2}L_{J}^{2}}{R_{l}}, & {\rm for}\;\omega \ll \omega _{LC} \\
R_{l}, & {\rm for}\;\omega =\omega _{LC} \\
\frac{1}{\omega ^{2}C_{sh}^{2}R_{l}}, & {\rm for}\;\omega \gg \omega _{LC}
\end{array}
\end{equation}
For the circuit with the $RC$ shunt
\begin{equation}
{\rm Re} \{Z_{t}(\omega)\}\approx
\begin{array}{cc}
\frac{\omega ^{2}L_{J}^{2}}{R_{l}}, & {\rm for}\;\omega \ll \omega
_{LC}\;\;\;\;\;\;\;\;\;\; \\
\lesssim R_{l}, & {\rm for}\;\omega =\omega _{LC}\ll \frac{1}{R_{sh}C_{sh}}
\\
R_{l}//R_{sh}, & {\rm for}\;\omega =\omega _{LC}\gg \frac{1}{R_{sh}C_{sh}}
\\
R_{l}//R_{sh}, & {\rm for}\;\omega \gg \omega _{LC}\;\;\;\;\;\;\;\;\;\;
\end{array}
\end{equation}
The difference mainly concerns frequencies $\omega >\omega _{LC}$, where the
$C$-shunted circuit has a ${\rm Re} \{Z_{t}(\omega )\},$ and thereby a
relaxation rate, that is several orders lower than for the $RC$-shunted
circuit.

For a $C$-shunted circuit with $\omega _{LC}\ll \omega_{res}$ the ${\rm Re}
\{Z_{t}(\omega _{res})\}\approx \frac{1}{\omega _{res}^{2}C_{sh}^{2}R_{l}}$.
This yields for $J(\omega)$ at $\omega > \omega_{LC}$
\begin{equation}
J(\omega)\approx \frac{\left( 2\pi \right) ^{2}}{\hbar }
\frac{1}{\omega^{3}}\left( \frac{MI_{p}}{\Phi _{0}}\right) ^{2}I_{sq}^{2}\tan
^{2}f\;\frac{1}{C_{sh}^{2}R_{l}}
\end{equation}
The factor $1/\omega ^{3}$ indicates a natural cut-off for
$J(\omega) $, which prevents the ultraviolet divergence 
\cite{leggett87,grifoni99} and which in much of the theoretical 
literature is introduced by hand. 
The $RC$-shunted circuit has softer
cut off $1/\omega$. The mixing rate for the $C$-shunted circuit is
then
\begin{equation}
\begin{array}{c}
\Gamma _{r}\approx \frac{\left( \Delta /\hbar \right) ^{2}}{\omega _{res}^{5}}
\frac{\left( 2\pi \right) ^{2}}{2\hbar }\left( \frac{MI_{p}}{\Phi _{0}}\right) ^{2}
I_{sq}^{2}\tan ^{2}f\;  \times \\
\frac{1}{C_{sh}^{2}R_{l}}\coth \left(
\frac{\hbar \omega _{res}}{2k_{B}T}\right) .
\end{array}
\end{equation}
Fig.~\ref{figtr} presents mixing times $\tau _{r}$ vs $\omega _{res}$ for
typical sample parameters (here calculated with the non-approximated version
of ${\rm Re} \{Z_{t}(\omega)\}$). With the $C$-shunted circuit it seems possible to get
$\tau _{r}$ values that are very long. They are compatible with the ramp
times of the SQUID, but too slow for fast repetition rates.
In Fig.~\ref{figsqrez} one can directly see from
the values of $J(\omega )$ that an
$RC$-shunted circuit with otherwise similar parameters yields at $\omega
_{res}/2\pi =$ 10 GHz relaxation times that are about four orders shorter.
For the parameters used here they are in the range of 15 ${\rm \mu s}$. 
While this value is close to the desired order of magnitude, 
one has to be aware of the fact at these high switching current values
the linearization Eq.\ (\ref{Lj}) may underestimate the actual noise. 
In that regime, phase diffusion between different minima of the 
washboard potential also becomes relevant and changes the noise properties
\cite{joyez99,coffey96}. 

\subsection{Dephasing times}

\begin{figure}[htb]
\begin{minipage}{0.9\columnwidth}
\includegraphics[width=0.8\columnwidth]{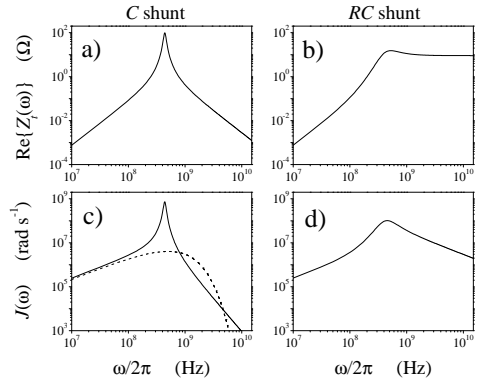}
\caption{A typical ${\rm Re}\{Z_{t}(\protect\omega)\}$ for the
$C$-shunted SQUID (a) and the $RC$-shunted SQUID (b), and
corresponding $J(\protect \omega)$ in (c) and (d) respectively.
For comparison, the dashed line in (c) shows a simple Ohmic
spectrum (\ref{JwOhm}) with exponential cut off $\protect\omega_c
/ 2\protect\pi$ = 0.5 GHz and $\protect\alpha$ = 0.00062. The
parameters used here are $I_{p} = 500$ nA and $T = 30$ mK. The
SQUID with $2I_{co}$ = 200 nA is operated at $f = 0.75 \,
\protect\pi $ and current biased at 120 nA, a typical value for
switching of the $C$-shunted circuit (the $RC$-shunted circuit
switches at higher current values). The mutual inductance $M$ = 8
pH (i.~e.~$\frac{MI_{p}}{\Phi_{0}} = 0.002$). The shunt is
$C_{sh}$ = 30 pF and for the $RC$ shunt $R_{sh} = $ 10 $\Omega $.
The leads are modeled by $R_{l} = 100$ $\Omega $.}
\label{figsqrez}
\end{minipage}
\end{figure}

\begin{figure}[htb]
\begin{minipage}{0.9\columnwidth}
\includegraphics[width=0.8\columnwidth]{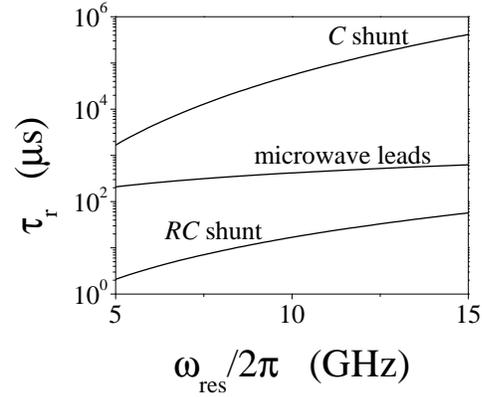}
\caption{Typical relaxation times due to the $C$-shunted SQUID,
the $RC$-shunted SQUID, and coupling the microwave leads as a
function of the resonance frequency at which the qubit is
operated. The example of the microwave leads contribution is for a
mutual inductance $M_{mw}$ to the coaxial line of $M_{mw} = 0.1$
pH. Parameters are further as described in the caption of
Fig.~\ref{figsqrez}.} \label{figtr}
\end{minipage}
\end{figure}

\begin{figure}[htb]
\begin{minipage}{0.9\columnwidth}
\includegraphics[width=0.8\columnwidth]{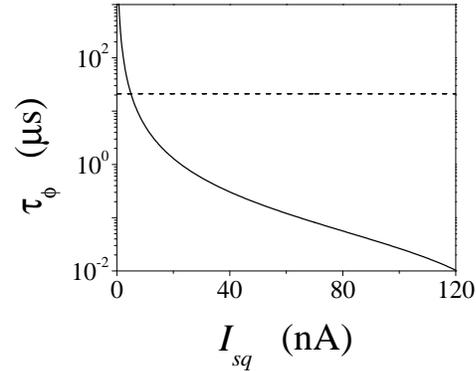}
\caption{The dephasing time (\ref{GphaseSQ}) as a function of the
bias current $I_{sq}$ through the SQUID (solid line). 
The dashed line shows $\protect\tau_{\protect\phi}$ for
$I_{sq} = 5$ nA, a typical minimum value for the effective bias
current for a SQUID with a few percent asymmetry between its
junctions. At this point $\protect\tau_{\protect\phi} = 131$
$\protect\mu$s, and $\protect\alpha = 1\cdot 10^{-7}$. Parameters
are further as described in the caption of Fig.~\ref{figsqrez}. }
\label{figtphi}
\end{minipage}
\end{figure}

At low frequencies $\omega <\omega _{LC}$ the $C$-shunted and $RC$-shunted
scheme have ${\rm Re} \{Z_{t}(\omega )\}\approx \frac{\omega ^{2}L_{J}^{2}}
{R_{l}}$ such that with (\ref{Jwsquid}) and (\ref{alpha})
\begin{equation}
J(\omega )\approx \frac{\left( 2\pi \right) ^{2}}{\hbar }\omega
\left( \frac{MI_{p}}{\Phi _{0}}\right) ^{2}
I_{sq}^{2}\tan ^{2}f\;\frac{L_{J}^{2}}{R_{l}},
\end{equation}
\begin{equation}
\alpha \approx \frac{2\pi}{\hbar }
\left( \frac{MI_{p}}{\Phi _{0}}\right) ^{2}
I_{sq}^{2}\tan ^{2}f\;\frac{L_{J}^{2}}{R_{l}}.
\label{alphaLR}
\end{equation}
The environment is Ohmic at low frequencies since we have
$J(\omega )\propto \omega $. For
our sample parameters the second term in (\ref{Gphase}) dominates, such that
with (\ref{alphaLR}) and for the qubit operated where $\varepsilon \approx \nu $
\begin{equation}
\Gamma _{\phi }\approx \frac{\left( 2\pi \right) ^{2}}{\hbar ^{2}}\left(
\frac{MI_{p}}{\Phi _{0}}\right) ^{2}I_{sq}^{2}\tan ^{2}f\;
\frac{L_{J}^{2}}{R_{l}}\,k_{B}T.
\label{GphaseSQ}
\end{equation}
Note in (\ref{Lj}) that $L_{J} \propto 1/I_{co}$, such that the dephasing
rate (\ref{GphaseSQ}) does not depend on the absolute value of the current, but
on the ratio $I_{sq}/I_{C}$.
For the typical sample parameters as used in Fig.~\ref{figsqrez}
the dephasing time is about 10 ns, which is too short.
However, we can gain a few
orders (if $\Gamma_{r}$ is low enough) by the fact that we can do the
quantum coherent control at low $I_{sq}$ (the previous estimate was calculated
for $I_{sq}=120$ nA, in the switching region). At $I_{sq}=0$ we
find $\Gamma _{\phi }=0$ in this linear approximation for the SQUID
inductance. At $I_{sq}=0$ we should therefore estimate the dephasing due to
second order terms. However, in practice the
dephasing is probably dominated by the second term in (\ref{circurnoise2}),
which is due to a small asymmetry in the fabricated SQUID junctions of a few
percent. This influence can be mapped on a small bias current (a few percent
of the critical current, say 5 nA) through the SQUID. Therefore, at
$I_{sq}\approx 0 $ the dephasing times can be $\left( \frac{120}{5}\right)
^{2}$ times longer. Furthermore, the factor $L_{J}^{2}$, as defined in (\ref
{Lj}), is at 5 nA about a factor 2 lower than at 120 nA. For our parameters
this allows for $\tau _{\phi }\approx 20$ ${\rm \mu s}$, see also Fig.~\ref
{figtphi}. Further improvement is feasible by
making e.\ g.\ $R_{l}$ = 1 k$\Omega $, working with a
lower mutual inductance $M$ or tuning the qubit close to the degeneracy point, 
as in \cite{saclay02}.

Finally, we would like to mention that in the literature on dissipative 
two-level systems one often assumes Ohmic
dissipation, corresponding to a purely resistive shunt across the junctions
of the qubit. For a description of such a system one usually introduces
an artificial exponential cut off at frequency $\omega _{c}$,
yielding $J(\omega )$ of the form
\begin{equation}
J(\omega )= \alpha \omega \exp (-\frac{\omega }{\omega _{c}}).
\label{JwOhm}
\end{equation}
In our case, $J(\omega)$ has substantial internal structure originating from
the frequency-dependence of ${\rm Re} \{Z_{t}(\omega )\}$. In order to compare our
results to the Ohmic case,
we plot in
Fig.~\ref{figsqrez} an Ohmic fit to the actual $J(\omega )$ of the
DC SQUID. For the parameters as in
Fig.~\ref{figsqrez} the resemblance is reasonable for a resistive
shunt corresponding to $\alpha =0.00062$, and a cut off $\omega
_{c}=0.5$ GHz.  For low currents, as for the dashed line in
Fig.~\ref{figtphi} $\alpha=1 \cdot 10^{-7}$. This corresponds to
an extremely underdamped system, with a long dephasing time.


\section{Relaxation and dephasing from on-chip control circuits}

\begin{figure}[htb]
\begin{minipage}{0.9\columnwidth}
\centerline{\includegraphics[width=0.6\columnwidth]{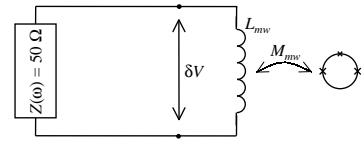}}
\caption{Circuit model for coaxial line that is inductively
coupled to the qubit. The coaxial line is modeled as a 50 $\Omega$
impedance that is shorted near the qubit with an inductance
$L_{mw}$. The qubit is coupled to this short with a mutual
inductance $M_{mw}$. The noise that couples to the qubit results
from Johnson-Nyquist voltage noise $\protect\delta V$ from the
circuit's total impedance $Z_t$, formed by a parallel combination
of the 50 $\Omega$ impedance and $L_{mw}$.} \label{figmwcir}
\end{minipage}
\end{figure}

We will treat here the influence of noise from the microwave leads in a
similar way as worked out for the SQUID. Here the total environmental
impedance $Z_{t}(\omega)$ is formed by a $50\;\Omega $ coax, that is shorted at the
end by a small inductance $L_{mw}$, see the circuit model in Fig.~\ref
{figmwcir}. This inductance $L_{mw}$ has a mutual inductance $M_{mw}$ to the
qubit. The voltage noise is given by (\ref{JNnoise}). The noise leads to
fluctuations $\delta \varepsilon $ of the energy bias separation
$\varepsilon $ as follows. The current noise in $L_{mw}$ is $\delta I_{L}=
\frac{1}{i\omega L_{mw}}\delta V$. The qubit fluctuations in the flux
$\Phi_{q}$ are then
$\delta \Phi _{q}=M_{mw}\delta I_{L}$, and the fluctuations
in the energy bias are $\delta \varepsilon =2I_{p}\delta \Phi _{q}$.
This gives for the fluctuations $\langle \delta \varepsilon \;\delta
\varepsilon \rangle _{\omega }$
\begin{equation}
\langle \delta \varepsilon \;\delta \varepsilon \rangle _{\omega }=
\frac{4\hbar }{\omega }\left( \frac{M_{mw}I_{p}}{L_{mw}}\right) ^{2} {\rm Re}
\{Z_{t}(\omega )\}\coth \left( \frac{\hbar \omega }{2k_{B}T}\right)
\end{equation}
and for the environmental spectral density (\ref{Jw}) 
\begin{equation}
J(\omega )=\frac{4}{\hbar \omega }\left( \frac{M_{mw}I_{p}}{L_{mw}}\right)
^{2} {\rm Re} \{Z_{t}(\omega )\}.
\end{equation}
The ${\rm Re} \{Z_{t}(\omega )\}$ is that of a first order low-pass $LR$
filter with a -3 dB frequency $\omega _{LR}=R/L$. For the $L_{mw}$ to be
effectively a short its impedance $\omega L_{mw}$ should be small compared
to 50 $\Omega $ at the frequency of the applied microwave radiation
(typically 10 GHz), giving $L_{mw}\ll 1\,{\rm nH}$. This can be realized by
making the length of the short line less than about 100 ${\rm \mu} $m.
This means
that all relevant frequencies are below the -3 dB frequency $\omega _{LR}$,
and that for both relaxation and dephasing we can approximate
\begin{equation}
{\rm Re} \{Z_{t}(\omega )\}\approx \frac{\omega ^{2}L_{mw}^{2}}{R_{mw}}
\end{equation}
with $R_{mw}=50\,\Omega $. For $\omega <\omega _{LR}$, $J(\omega )$ is again
Ohmic,
\begin{equation}
J(\omega )=\frac{4\omega }{\hbar }\frac{\left( M_{mw}I_{p}\right) ^{2}}
{R_{mw}},
\end{equation}
and with (\ref{alpha}) we find for $\alpha $
\begin{equation}
\alpha \approx \frac{4}{2 \pi \hbar }\frac{\left( M_{mw}I_{p}\right) ^{2}}{R_{mw}}.
\end{equation}
Note that these results are independent of $L_{mw}$. A larger $L_{mw}$ leads
to enhanced voltage noise, but the resulting current noise is reduced by the
same factor. For frequencies below $\omega _{LR}$ the current noise is just
that of a shorted 50 $\Omega $ resistor. For frequencies higher than $\omega
_{LR}$, ${\rm Re} \{ Z_{t}(\omega ) \} \approx R_{mw}$, such that $J(\omega )
$ has a very soft intrinsic $1/\omega $ cut off.

For the relaxation rate (\ref{Gmix})\ as a function of the qubit's resonance
frequency $\omega _{res}$ we now have
\begin{equation}
\Gamma _{r}\approx \frac{2\Delta ^{2}}{\hbar^{3} \omega _{res}}\frac{\left(
M_{mw}I_{p}\right) ^{2}}{R_{mw}}\coth \left( \frac{\hbar \omega _{res}}
{2k_{B}T}\right) .
\end{equation}
This has a much weaker dependence on $\omega _{res}$ than for the SQUID,
results are plotted in Fig.~\ref{figtr}. The results are plotted for
$M_{mw}=0.1\,{\rm pH}$ (further parameters are as used for the SQUID
calculations) and for this $M_{mw}$ value the relaxation times are in the
required range of about 100 ${\rm \mu s}$. The value $M_{mw}\approx 0.1\,
{\rm pH}$ corresponds to a 5 ${\rm \mu m}$ loop at about 25 ${\rm \mu m}$
distance from the microwave line, and is compatible with the fabrication
possibilities and the microwave requirements. With this geometry it is still
possible to apply sufficient  microwave power for pumping the qubit's Rabi
dynamics at 100 MHz (i.~e.~pumping with an oscillating $\Phi _{q}$ of about
0.001 $\Phi _{0}$ \cite{mooij99}, 
which needs an oscillating current of $\frac{0.001\,\Phi _{0}}
{M_{mw}}\approx 20\,{\rm \mu A}$, corresponding to $20\,{\rm nW}$, i.~e.~
$-47\,{\rm dBm}$
microwave power), while the dissipated microwave power in the attenuators at
the refrigerator's base temperature remains well below the typical cooling power 
of 1 ${\rm \mu W}$.

For the second term of the dephasing rate (\ref{Gphase})\ we thus find for
the qubit operated where $\varepsilon \approx \nu
$\begin{equation}
\Gamma _{\phi }\approx \frac{4}{\hbar ^{2}}\frac{\left(
M_{mw}I_{p}\right) ^{2}}{R_{mw}}\,k_{B}T.
\end{equation}
Using the same parameters as
in the above calculation of the relaxation time we find $\alpha \approx
1\cdot 10^{-7}$ and for $T=30\,{\rm mK}$ the dephasing time is
$\tau_{\phi }\approx 130\,{\rm \mu s}$.
While this dephasing rate is sufficient for demonstration experiments
and promising for applications, we like to note that it is much harder to engineer 
this dephasing rate as compared to the DC-SQUID dephasing.
It is in practice quite tedious to apply microwave technology
with impedance levels $R_{mw}$ much higher than 50 $\Omega $, both for externally
generated microwaves and on chip generators (it could for instance be 
increased using a planar impedance transformer \cite{feldman2001}). 
Making $M_{mw}$ smaller requires higher microwave
currents, and thereby more microwave dissipation on the mixing chamber. The
cooling power per qubit will quite likely remain below 1 ${\rm \mu W}$, so
much stronger microwave signals from a larger distance is not an option.
Moreover, making $M_{mw}$ very small means that the control line is 
$100 {\rm \mu m}$ or further away from the
qubit. In this case, applying microwaves locally
to one specific qubit on a chip with several coupled qubits is much harder.


\section{Suppressing rates by freezing states and
idle states}

With additional control techniques the decoherence rates can be 
made better than the estimates made in the previous sections. 
These are based on the pre-factors in Eqs. (\ref{Gmix}) and
(\ref{Gphase}).
Bringing the qubit in a so-called idle state ($\varepsilon=0$) \cite{makhlin99}
can reduce the dephasing, but not beyond $\Gamma_{r}/2$
(which is enhanced at $\varepsilon=0$).

\begin{figure}[htb]
\begin{minipage}{0.9\columnwidth}
\includegraphics[width=0.8\columnwidth]{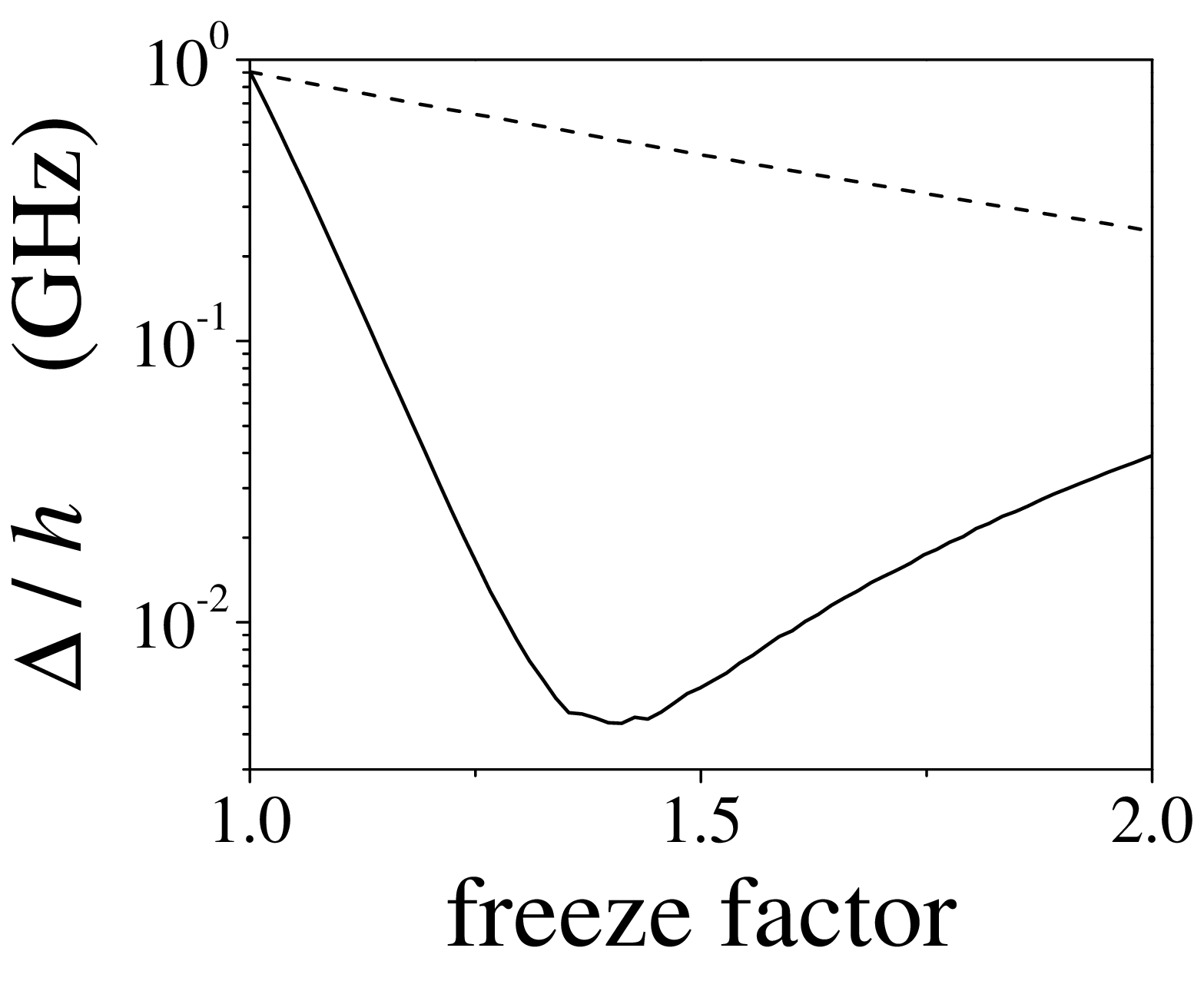}
\caption{Numerically simulated suppression of $\Delta$ in qubits where one (solid) or
three (dashed) of the junctions are realized as small DC-SQUIDs, 
using typical qubit parameters as mentioned in the text.
The horizontal axis is the factor by which the Josephson energy of
the relevant junctions is increased. For the solid line, 
one can observe a non-monotonic dependence of $\Delta$ on the
freeze factor, which results from the fact that if the Josephson coupling of 
the weak junction is strongly increased, alternative tunneling paths between
the current states open up, see also Ref.\ \cite{mooij99}.} 
\label{figfreez}
\end{minipage}
\end{figure}

Another useful technique is freezing. 
Here, the tunnel coupling $\Delta$ between the two qubit states is
strongly reduced before the measurement process starts
with a fast but adiabatic control current \cite{mooij99}.
This allows for much slower measurements, thus
for weaker coupling to a damped SQUID with very high resolution.
Moreover, it has the advantage that the tunnel coupling becomes
so week that the Hamiltonian almost commutes with $\hat{\sigma}_{z}$.
This can improve the correlation between the outcome of $\hat{\sigma}_{z}$
measurements, and energy states in the case that the calculation
states are energy eigen states.
However, freezing requires that the qubit junctions
are realized as small DC-SQUIDs. This means that $\hat{\sigma}_{x}$
noise will be strongly enhanced. The influence of this additional
noise source can be calculated along the same lines as in \cite{grifoni99}
and, assuming that it is not correlated with the other noise sources, 
add up to the rates we have calculated in this paper.
See Fig.~\ref{figfreez} for numerical estimates
on the possible adiabatic suppression of $\Delta$.

\bigskip


\section{Discussion and conclusion}

We have developed a scheme for modeling the decoherence of a persistent-current
qubit due to its electromagnetic environment. Examples for both control and read-out 
electronics are worked out quantitatively. We discussed how the dephasing and 
relaxation rates of the qubit can be derived from the impedance
of the electronic circuitry, and provided design criteria for such electronics.
In particular, we have shown that even though the readout SQUID is always close to the 
qubit, it can be effectively decoupled. 
Our examples show that the present status of experimental technology
should allow for the observation of quantum coherent oscillations 
between macroscopic persistent-current states.

In this final section, we like to point out that the theory that is used is still in development. 
In particular, an environment with a strongly structured spectrum may violate the 
weak-coupling Born approximation at its resonance frequencies, but it may
also induce weak additional decoherence off-resonance. This
situation is hardly accessible with traditional theoretical methods for
this problem and alternative approaches such as the flow-equation scheme 
in \cite{Kleff} may be needed. Furthermore, it is not clear
whether the impedance can be described by a single temperature.
At low frequencies, noise from parts in the system with a higher
temperature can reach the sample. However, the experimental results 
reported in Ref. \cite{saclay02} indicate, that 
an analysis of the decoherence of the type we give here gives good 
predictions of the experimental decoherence time scales in a superconducting
qubit.

\bigskip

The authors thank D. Esteve, M. Grifoni, Y. Nakamura, P. C. E.
Stamp, A. C. J. ter Haar, L. Levitov, T. P. Orlando, L. Tian, and
S. Lloyd for help and stimulating discussions. Financial support
by the Dutch Foundation for Fundamental Research on Matter (FOM),
the European TMR research network on superconducting nanocircuits
(SUPNAN), the USA National Security Agency (NSA) and Advanced
Research and Development Activity (ARDA) under Army Research
Office (ARO) contract numbers DAAG55-98-1-0369 and P-43385-PH-QC, 
and the NEDO joint
research program (NTDP-98) is acknowledged.


\pagebreak

\begin{thebibliography}{}
\bibitem{manyloop} Alternatively, one could also take an 
NMR-like approach and measure on an ensemble of loops. This
is not discussed in this article.
\end{thebibliography}
\end{document}